\documentclass[twocolumn,aps,prl,superscriptaddress,groupedaddress]{revtex4}
\usepackage{comment}
\usepackage{color}
\usepackage{amsmath}
\usepackage{amssymb}
\usepackage{graphicx}
\usepackage{esint}
\usepackage{adjustbox}
\usepackage{comment}
\usepackage{braket}
\usepackage[unicode=true,pdfusetitle,
bookmarks=true,bookmarksnumbered=false,bookmarksopen=false,
breaklinks=true,pdfborder={0 0 1},backref=false,colorlinks=true]
{hyperref}
\hypersetup{
	linkcolor=red,urlcolor=blue,citecolor=blue,anchorcolor=blue}
\usepackage{breakurl}

\makeatletter
\@ifundefined{textcolor}{}
{%
	\definecolor{BLACK}{gray}{0}
	\definecolor{WHITE}{gray}{1}
	\definecolor{RED}{rgb}{1,0,0}
	\definecolor{GREEN}{rgb}{0,1,0}
	\definecolor{BLUE}{rgb}{0,0,1}
	\definecolor{CYAN}{cmyk}{1,0,0,0}
	\definecolor{MAGENTA}{cmyk}{0,1,0,0}
	\definecolor{YELLOW}{cmyk}{0,0,1,0}
}

\usepackage{dcolumn}\usepackage{bm}

\makeatother

\begin{document}
	
	\title{CAVITY RING-DOWN SPECTROSCOPY WITH BEHAVIOR OF HYBRID CAVITY STRUCTURES}
	
	\author{Muhammad Junaid Khan}
	
	\email{mjkhan@ele.qau.edu.pk}

	\affiliation{Department of Physics, COMSATS University Islamabad, Islamabad Pakistan}
	
	\author{Rida Batool Sheraliyat}
    \affiliation{Department of Electronics, Quaid-i-Azam University, Islamabad, Pakistan}
	\affiliation{Department of Physics, COMSATS University Islamabad, Islamabad Pakistan}

	\date{\today}

\begin{abstract}
Cavity ring-down (CRD) spectroscopy represents a direct absorption technique of sample absorption measurement. Instead of measuring the amount of the absorbed light, this technique determines the rate at which light intensity decays inside an optical cavity. When using a pulsed or continuous-wave light source, CRD spectroscopy offers considerably higher sensitivity as compared with conventional spectroscopy of absorption, making even very weak absorptions easy to detect.
	\end{abstract} 
    
\maketitle

\section{Introduction} \label{sec1}
Cavity ring-down spectroscopy (CRDS) was introduced in the 1980s by O’Keefe et al. \cite{o1988cavity}. This technique makes it feasible to measure absorption using either a pulsating or a continuous wave of light sources\cite{romanini1997cw}. 

As we note, this article will emphasize Cavity Ring-Down Spectroscopy (CRDS). As with so many scientific discoveries, CRDS began as an outgrowth of large advancements in technological capabilities, namely that of greatly enhancing the reflectivity of high-quality dielectric mirrors \cite{bilger1994origins}. While the transmission factor \(T\) into these mirrors can be measured relatively sensitively, there are additional sources of light loss caused by absorption and scattering of light within the materials composing the coatings on dielectric mirrors. It soon became realized that a stable optical resonator could be formed with two mirrors of this sort, and that it took a characteristic time inversely proportional to total loss per reflected pulse and, more precisely, proportional to $1-R$, reflecting inverse proportionality to mirror reflectivity $R$. Higher values correspond to larger photon storage times within the cavity.

A major contribution toward the development of CRDS was made by Anderson, Frisch, and Masser \cite{anderson1984mirror}. They injected a laser pulse into an optical cavity, thus allowing the cavity to be energized, and then monitored the decay of the intracavity light leaking out through one of the mirrors. The intensity inside the cavity decays exponentially with time. The decay rate k is given by

\begin{equation}
   k = \frac{(1 - R)c}{L} 
\end{equation}

where $c$ is the speed of light, $L$ is the length of the cavity, and $R$ is the reflectivity of the mirrors. The inverse of $k$, $1/k$, is the ring-down time $\tau$ and characterizes the average time a photon spends inside the cavity.

The effective optical path length is

\begin{equation}
 L_{\text{eff}} = \tau c = \frac{L}{1 - R}   
\end{equation}

which corresponds to the total distance a photon travels inside the optical cavity before it escapes or is absorbed.

A pair of mirrors with a separation distance $L=4.0mm$ was employed by Rempe et al. \cite{rempe1992measurement}. When the cavity was driven at a wavelength $\lambda=850nm$, they measured a loss of $1.6ppm$ due to each mirror.

The experiment done by Rempe et al. played a significant role in the development of CRDS, since it provided an experimental demonstration that ultra-high-reflectivity dielectric mirrors could be fabricated with reliably low loss. A high-finesse cavity was hence enabled for the construction of not only CRD absorption measurements but also a wide range of advanced optical research areas that comprises cavity quantum electrodynamics, optical frequency standards, quantum memories, and precision metrology. Their demonstration of ppm-level mirror losses remains to date one of the basic references in modern CRDS system design and other optical instruments at high sensitivity.
 
 \par
\section{Experimental set-up}\label{sec2}
\begin{figure}[h!]
    \centering
    \includegraphics[width=9cm,height=8cm]{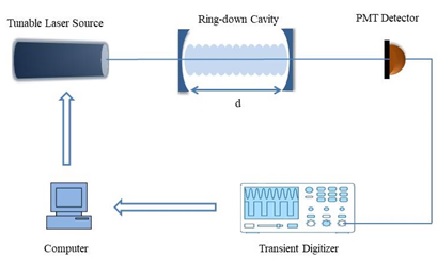}
    \caption{Schematic diagram of a Cavity Ring-down Spectroscopy. The absorption cell with length d. We use nitrogen laser Pumping providing energy to dye laser. A 10 nanosecond laser pulse from a dye laser which powered by a Nitrogen laser.}
    \label{CRDSfig1}
\end{figure}

A high-finesse Fabry-Pérot cavity is created with two highly reflective mirrors, making it possible to trap pulsed lasers with variable wavelengths. The experimental arrangement for cavity ring-down spectroscopy is shown in Figure \ref{CRDSfig1}. It should be noted that a ring-down spectroscopy experiment requires a relatively simple configuration as it consists of a tunable pulsed laser source, a ring-down cavity, a fast photodetector, a transient digitizer, and a computer. As ring-down spectroscopy is an absorption spectroscopy technique conducted at various wavelengths, it requires precise tuning of laser wavelengths. Based on visible wavelengths, ring-down spectroscopy experiments have highly sensitive capabilities and are capable of measuring very low absorbance signals at low pulse energy with pulse duration ranging from 5-15 ns. The energy for these laser pulses can be measured as approximately 1 $mJ$. As alternative sources for pulsed ultraviolet light, Raman-shifting techniques have been employed on visible laser light sources \cite{grossmann1987raman}. When conducted at visible regions, ring-down spectroscopy experiments have mirror reflectivities on the order of $1-R\sim 10^2-10^3$. At these levels, it becomes more feasible to employ a photomultiplier tube as experimental detectors due to their high sensitivity \cite{thompson2006cavity}.

Here, the Fabry–Pérot cavity acts as an optical storage system where the injected laser pulse is forced to make several thousand round trips between the two mirrors while its intensity decays. Since the reflectivity of the mirrors is extremely high, only a minuscule proportion of the light escapes upon every pass. This “ring-down” signal, representing the gradual decay of the trapped light, is monitored by a detector positioned behind one of the mirrors \cite{engeln1997polarization, hodges2004frequency, lisak2022dual}. Despite the fact that the mirror is very highly reflective, it transmits a small but well-defined portion of the circulating light, allowing measurement of the decay curve with high accuracy.

Once the laser pulse is coupled into the cavity, no additional input light is permitted to enter, and the subsequent decay represents a purely passive measurement that is dependent only on the intrinsic losses inside the cavity. The losses in the cavity include mirror transmission, scattering, diffraction, and most importantly, absorption by the sample placed inside the cavity \cite{guan2022resonant}. Any absorbing species (whether gas, aerosol, or thin film) increases the loss inside the cavity by a factor that is very small. This additional cavity loss shortens the ring-down time, and by comparing the decay constants with and without the sample, extremely small absorption coefficients can be quantified with high accuracy.

The exponential decay signal is acquired by the transient digitizer with nanosecond resolution, in order for the ring-down time constant to be extracted by fitting the measured output to an exponential function. Since this is a time-of-decay measurement rather than an absolute intensity measurement, this technique has a high degree of immunity to fluctuations in laser power, beam pointing, or pulse-to-pulse variations. The immunity to intensity noise is another one of the major reasons why CRD spectroscopy achieves superior detection limits compared to conventional absorption spectroscopy \cite{crosson2008cavity}.

When the laser wavelength is scanned through the absorption line, the ring-down time will change. Wavelengths coinciding with molecular transitions will have a shorter decay time because of enhanced absorption, whereas off-resonant wavelengths retain the longer baseline decay time, which is determined mainly by mirror losses. By recording the ring-down time as a function of wavelength, high-resolution absorption spectra can be retrieved, even for species present at extremely low concentrations.

For experiments in the visible region, where mirror reflectivities normally give losses of the order of $10^{-2}$ to $10{-3}$ per round trip, the photomultiplier tubes are well fitted because they have fast temporal response and are capable of detecting very weak light levels. The PMT converts these leaked photons into electrical pulses, which are then digitized and analyzed \cite{fitzpatrick2013single}. In the more advanced setups of sub-nanosecond pulses, ultraviolet wavelengths, or cavities with extremely low loss, other detectors may be used, including avalanche photodiodes and fast photodiodes, but PMTs remain standard instruments owing to their excellent sensitivity and speed. Overall, a high-finesse cavity, rapidly tunable pulsed lasers, and fast detection electronics together enable CRD spectroscopy to measure absorption coefficients as low as $10^{-9}–10^{-10}$ $cm^{-1}$. This makes the method suitable for applications such as trace-gas detection, kinetic studies, and precise measurements of optical constants. If the cavity is further stabilized or coupled with narrowband continuous-wave lasers, the technique can be extended to achieve even higher sensitivity and spectral resolution.

During the wavelength scan, it is necessary that the laser frequency be stable with respect to the cavity modes \cite{thorpe2006broadband}. A Fabry-Pérot cavity permits only specific resonant modes depending on its length. Only laser pulses with wavelengths matched with an integral number of cavity lengths will be coupled effectively into the cavity. Any mismatch within a very small margin will result in a large attenuation of the injection efficiency, and as a consequence, no buildup will occur within the cavity. To maintain a stable injection efficiency throughout a wavelength scan, it is common practice for CRD systems to include either an automatic cavity length controller or a laser frequency locking loop, with which the laser frequency will be automatically locked on the closest cavity mode \cite{engeln1997polarization}.

Once the laser pulse has been successfully entered into the cavity, a simple exponential decay will occur for the trapped field, but several processes will be taking place at once. A small amount of loss will occur with each round trip within the cavity. The decay will be a result of the total loss per round trip, which will be a result of mirror transmission, mirror absorption, scattering from imperfections, diffraction loss due to mirror aperture size, and sample absorption or scattering \cite{vainio2016mid}. The ring-down time, a measure of time for the field intensity to decrease by 1/e from its initial value, will be extremely sensitive to even nominal changes within the total loss. As a result, if there is a nominal additional absorption coefficient introduced into the sample, there will be a measurable ring-down time, and as such, CRD spectroscopy will be highly sensitive compared with other optical absorption spectroscopies.

During experiments, these digitized decay signals are averaged multiple times depending on the laser pulses. The laser sources exhibit fluctuations from pulse to pulse in terms of energy and timing jitter. Despite these fluctuations among laser pulses, because CRD experiments do not make use of the peak intensities, they do not affect the measured decay times. The laser sources effectively make all these pulse sources have the same exponential decay rate due to losses \cite{long2012frequency}.

The computer system is an essential component of an analysis process, as it uses an exponential fit on a measured decay trace. A perfectly aligned cavity with as few sources of noise as possible will have a single exponential decay. It should be noted that some imperfections might occur with realistic ring-down measurements. Misalignment could result in a cavity that supports a higher transverse mode, which would have a slightly differing decay. Also, stray reflections, poorly triggered signals, and some electrical noise might affect either the early or late regions on a ring-down trace. Once the ring-down times have been determined at various wavelengths, it becomes possible to calculate directly the absorption spectrum. The difference between the ring-down times for the pure and sample-filled cavities becomes a highly precise measurement of the sample's absorption. Because it avoids absolute calibration, which generally represents the source of error for conventional absorption spectroscopy, CRD spectroscopy becomes capable of measuring very weak transitions, trace levels of gases, and sharp spectral features that might be below the detection limit with conventional methods. Due to its high sensitivity, it has been widely applicable to various domains, ranging from environmental observation and combustion analysis to atmospheric chemistry and optical materials research. The capability for trace absorption line detection also renders it highly advantageous for researching species with very low concentrations or transitions with very small oscillator strengths. Moreover, with varying cavity designs or employing mirrors with very high reflectivities, it is feasible to extend these same fundamental concepts into either the ultraviolet and near infrared or even mid-infrared regions.

Martyn D. Wheeler et al give a detailed explanation of their setup in reference \cite{wheeler1997predissociation}. In our setup, the ring-down cavity forms the central part of the system. Two high-reflectivity (HR), dielectric-coated concave mirrors enclose the absorption region. These mirrors normally have a maximum reflectivity value of about 99.9\%, thus enabling the confined light to experience multiple successive reflections. The ring-down cavity is formed by a pair of identical plano–concave mirrors inside which the laser pulse is confined and allowed to bounce back and forth many times.

The configuration used here employs mirrors with a moderately small radius of curvature, typically between -25 $cm$ and -1 $m$. Such a mirror separation can be denoted as $d$, and for the cavity to remain optically stable, it has to fulfill the well-known geometric stability conditions. In the case of this setup, the cavity operates within a nonconfocal regime expressed as 

\[
0 < d < r \quad \text{or} \quad r < d < 2r
,\]

with $r$ being the radius of curvature of the mirrors. Indeed, the separation between the mirrors is 3 $cm$ in the present setup and hence meets the stability requirement, thus guaranteeing efficient confinement of light between the two high-reflectivity surfaces \cite{ullah2021phase, cheema2012simultaneous, sharma2018comparison}.

Much longer cavities have also historically been built. For example, there was a report by LeGrand and Le Floch in 1990 of a ring-down cavity 270-$cm$ in length, showing the quite general geometries and lengths that have been used in the technique of CRD spectroscopy \cite{le1990sensitive}.

\section{Cavity modes}\label{sec3}
This picture conveys the idea that the process does not depend on the frequency of the laser light used to excite the cavity. This is not true, since it has to consider the modes of the cavity. The modes of a cavity can introduce several complexities. There might be molecular absorption lines that are so fine and are found between two cavity modes, which ultimately make those features absent in the obtained spectrum. There can also be oscillations in the ring-down due to the exciting of multiple cavity modes. These oscillations make it difficult to measure a precise value of the decay time $\tau$.
Experimentally, a continuum in this manner can be easily obtained. Firstly, no mode-matching optics should be employed when injecting light into the cavity. Secondly, for a stable, confocal cavity, it follows that the distance between the mirrors, denoted by d, should satisfy

\[
0 < d < r \quad \text{or} \quad r < d < 2r
,\]

with r being the curvature radius of the mirror. Finally, it should be ensured that the cavity size, in terms of the diameter of the mirror, is not too small, so that the modes are contained in the cavity with sufficiently small diffraction losses.

Moreover, it has to be ensured that all modes leaving the cavity are measured with similar efficiency, since transverse modes are larger in extent. A practical example of a cavity that satisfies both conditions would be a 50 $cm$ long cavity with a 25 $mm$ diameter for both mirrors and a curvature of radius $r=1$ $m$. In this situation, all modes can be measured by a photomultiplier placed immediately after the cavity. When a detector of small size is used, a lens of suitable diameter with a small focal length would be needed to focus the transmitted rays on it.

\section{Pulsed cavity ring-down spectroscopy}\label{sec4}
Martyn D. Wheeler et al. have given a detailed description of our experimental configuration. When the pulse duration of the laser is shorter than the round-trip time of light within the cavity \cite{wheeler1998cavity, zalicki1995cavity}, a train of separate pulses will be detected. These pulses correspond to portions of light that have traveled within the cavity and have been partially transmitted via one of the mirrors after each round-trip passage. The amplitude of these pulses will decrease progressively as a result of mirror losses and absorption within the sample. The distance between these pulses will correspond to the round-trip time for light within the cavity and can be described as follows:

\begin{equation}
    t_r = \frac{2L}{c}
\end{equation}

Here,  $L$  is the cavity length and c is the speed of light in vacuum. When short laser pulses are used, the resulting pulse train in cavity ring-down measurements forms the basis for analyzing the exponential decay of light intensity within the cavity. The intensity of the light with respect to time can be analyzed for its exponential decay, although each round trip of the light between the mirrors will decay the intensity further.

If a laser pulse with an initial intensity $I$ laser is injected into an optical cavity of length $L$, the light will be trapped to circulate between the two mirrors. The light intensity decays according to the absorption of species in the cavity. A quantitative expression can be provided for the detected intensity in terms of mirror reflectivities, transmission of the mirrors, sample absorption coefficient, and optical path length.

Let us begin with the expression for the intensity of the first detected optical pulse in CRDS. In the typical cavity, the two identical mirrors have the reflectivity $R$ and transmission $T$, with $R$ + $T$ = 1. The reflectivity R defines a fraction of incident light reflected back into the cavity, while the transmission $T$ describes the fraction of light escaping the cavity. The transmitted fraction is usually much smaller than 1-$R$, often only about 10\% to 50\% of that value. Every time the light pulse travels from one mirror to the other, part of its intensity will be absorbed by a sample inside the cavity. The intensity of the first detected optical pulse in CRDS can be estimated according to the Beer-Lambert law that describes the dependence of light absorption upon the properties of the absorbing species:

\begin{figure}[h!]
    \centering
    \includegraphics[width=9cm,height=8cm]{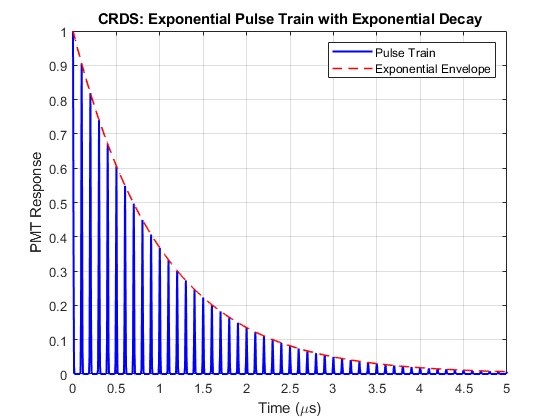}
    \caption{CRDS signal showing a pulse train (blue) with its exponential decay envelope (red). The decay rate of the envelope reflects the light loss in the cavity and is used to determine the sample’s absorption.}
    \label{CRDSfig2}
\end{figure}

\begin{equation}
   I = I_o e^{-\alpha d} 
\end{equation}

\begin{equation}
    I_o = I_{\text{laser}} T^2 e^{-\alpha d}
\end{equation}

The frequency-dependent absorption coefficient of the sample within the cavity will be represented as $\alpha(\nu)$. It will be assumed that the region with an extent of d contains the sample, and it will be within a cavity with an extent of L. Based on these considerations, the formula for the intensity within the second pulse within cavity ring-down spectroscopy will be given by:

\begin{equation}
    I_{\circ} = I_{\text{laser}} R^2 e^{-2 \alpha d}
\end{equation}

Since the mirrors used in cavity ring-down spectroscopy have a very high reflectivity $R \approx 1$, using the first-order Taylor expansion $\ln(R) \approx -(1-R)$, this can be rewritten as

\begin{equation}
    I(t) = I_o e^{\frac{t_c}{L} (\ln R - \alpha d)}
\end{equation}

\begin{equation}
    I(t) = I_o e^{\frac{t_c}{L} (1 - R + \alpha d)}
\end{equation}

\begin{equation}
    I(t) = I_0 \exp\left(-\frac{t}{\tau}\right)
\end{equation}

\section{Spectroscopy with polarized light}\label{sec2}
The rotationally resolved spectra of the

$\mathrm{b}^1\Sigma^+_g \; (\nu = 0) \leftarrow \mathrm{X}^3\Sigma^-_g \; (\nu = 0)$ band of molecular oxygen were studied in magnetic fields of up to 20 T in order to verify a theoretical model describing the interaction of oxygen molecules with the magnetic field. From this model, the spin contribution to the molar magnetic susceptibility of oxygen is obtained as a function of magnetic field strength and temperature. The results show that in strong magnetic fields, at relatively small temperatures, the molecular oxygen is aligned because of spin–spin interactions and because the spin angular momentum couples to the magnetic field: this proves that sensitive direct absorption spectroscopy can be performed inside short, high-field cavities where usually access is limited, such as inside a magnet.

In order to reduce the spectral broadening by magnetic-field inhomogeneity, we used a short ring-down cavity with length of just a few centimeters. It consisted of two plano-concave mirrors (25 mm diameter, -1m radius of curvature) separated by 3 cm and mounted on a fixed tube, which was inserted into a Bitter magnet with 3.1 cm bore, positioning the cavity along the magnetic field axis (Faraday configuration). We centered the cavity in the region of highest magnetic homogeneity, where the magnetic field value at the positions of the mirrors was reduced by only about 1.5\%, allowing for stable and high-resolution spectroscopy.

 \section{conclusion}\label{sec4}
Cavity Ring-Down Spectroscopy (CRDS) has been demonstrated as a highly effective direct absorption technique in gas-phase spectroscopy. It offers exceptional sensitivity, allowing the detection of molecules with very weak absorption signals as well as trace species exhibiting stronger absorption. CRDS is a robust and versatile method, capable of providing precise and accurate absorption measurements across a wide range of samples. Importantly, there are no inherent limitations that restrict CRDS to a specific spectral region, making it applicable from the ultraviolet to the infrared.

\bibliographystyle{apsrev4-1}
\bibliography{BEC.bib}

\end{document}